\documentclass[aps,prb,amsmath,amssymb,reprint,superscriptaddress,preprintnumbers,showpacs,intlimits]{revtex4-1}
\usepackage{bm,latexsym,mathrsfs,enumerate,color}
\usepackage[mathcal]{euscript}
\usepackage[breaklinks=true,unicode=true,urlcolor = blue,colorlinks = true,citecolor = blue,linkcolor = blue]{hyperref}
\usepackage{pgf,tikz}
\usepackage{graphicx}
\usepackage{todonotes}
\renewcommand{\vec}[1]{\bm{#1}}
\usepackage[normalem]{ulem}

\begin{document}
%
%
\title{Curvature induced domain wall pinning}

\author{Kostiantyn V. Yershov}
\email{yershov@bitp.kiev.ua}
\affiliation{Bogolyubov Institute for Theoretical Physics, 03143 Kyiv, Ukraine}
\affiliation{National University of ``Kyiv-Mohyla Academy", 04655 Kyiv, Ukraine}

\author{Volodymyr P. Kravchuk}
\email{vkravchuk@bitp.kiev.ua}
\affiliation{Bogolyubov Institute for Theoretical Physics, 03143 Kyiv, Ukraine}

\author{Denis D. Sheka}
\email{sheka@univ.net.ua}
\affiliation{Taras Shevchenko National University of Kyiv, 01601 Kyiv, Ukraine}

\author{Yuri Gaididei}
\email{ybg@bitp.kiev.ua}
\affiliation{Bogolyubov Institute for Theoretical Physics, 03143 Kyiv, Ukraine}

\date{\today}

%
%
%
%
\begin{abstract}
It is shown that a local bend of a nanowire is a source of pinning potential for a transversal head-to-head (tail-to-tail) domain wall. Eigenfrequency of the domain wall free oscillations at the pinning potential and the effective friction are determined as functions of the curvature and domain wall width. The pinning potential originates from the effective curvature induced Dzyaloshinsky-like term in the exchange energy. The theoretical results are verified by means of micromagnetic simulations for the case of parabolic shape of the wire bend. 
\end{abstract}

\maketitle
%
%
%
%
\section{Introduction}
Dynamics of domain walls in nanowires underlies modern memory\cite{Parkin08,Hayashi08} and logic\cite{Allwood02,Allwood05,Xu08a} devices, see also review Ref.~\onlinecite{Catalan12}. Most of these designs require wires with curvilinear segments, e.g. vertical configuration of the racetrack magnetic memory.\cite{Parkin08} Therefore, the modification of dynamic and static properties of a domain wall due to the wire curvature is of high importance for the applications. It is known that the local curvilinear defects of the wire can be a source of pinning potential for the domain wall.\cite{Lewis09,Glathe12,Burn14} However, in contrast to widely studied mechanisms of the domain wall pinning, such as interaction with artificial notches,\cite{Klaui03,Klaeui05,Hayashi06,Bedau07,Atkinson08,Bogart09,Petit10,Kim14} surface roughnesses,\cite{Nakatani03,Thiaville04,Thiaville05,Ivanov11} inhomogeneities in the magnetocrystalline anisotropy distribution,\cite{Uhlir10,Gerhardt11,Ivanov11} the role of the curvature remains unclear and the general theory is absent. Nevertheless, one should note some progress in studying the simplest case of circular wire geometry, where the domain wall pinning is achieved due to the applied external magnetic field.\cite{Kruger07a,Jamali11}

As it was recently shown \cite{Sheka15} the exchange interaction in a curvilinear wire can produce an effective Dzyaloshinsky-like term. In this paper we demonstrate that the inhomogeneous distribution of this curvature induced Dzyaloshinsky-like contribution can be an origin of domain wall pinning at the local wire bend, analogously to the case of studied anisotropy inhomogeneities.\cite{Uhlir10,Gerhardt11,Ivanov11} Considering the coordinate dependent  curvature of the wire we propose a general approach valid for a wide class of geometries. 

\section{Model and general results}
\label{sec:dw_dynamics}

In this paper we consider a plane curved ferromagnetic wire of circular cross-section. Such a wire can be parameterized in the following way:
\begin{equation} \label{eq:wire_definition}
	\vec{r}(s,\chi,\rho)=\vec{\gamma}(s)+\rho\cos\chi\vec{e}_\text{\sc{n}}(s)+\rho\sin\chi\vec{e}_\text{\sc{b}}(s).
\end{equation}
Here the three-dimensional radius vector $\vec{r}$ defines the space domain, occupied by the wire, the two-dimensional vector  $\vec{\gamma}(s)=\gamma_x(s)\hat{\vec x}+\gamma_y(s)\hat{\vec y}$ determines the wire central line, which lies within the $xy$-plane, with $s$ being the natural parameter (arc length). Within a wire cross-section the polar coordinates $\rho\in[0,\,R]$ and $\chi\in[0,\,2\pi)$ are used, where $R$ is the wire radius. We use here the Frenet-Serret basis ($\vec{e}_\text{\sc{t}},\,\vec{e}_\text{\sc{n}},\vec{e}_\text{\sc{b}}$) with $\vec{e}_\text{\sc{t}}=\vec{\gamma}'(s)$, $\vec{e}_\text{\sc{n}}=\vec{\gamma}''(s)/\kappa(s)$, and $\vec{e}_\text{\sc{b}}=\vec{e}_\text{\sc{t}}\times\vec{e}_\text{\sc{n}}$ being the tangential, normal, and binormal unit vectors, respectively. Here the wire curvature $\kappa(s)=|\vec{\gamma}''(s)|$ is introduced.

Using the Frenet-Serret basis one can introduce the angular magnetization parameterization 
\begin{equation}\label{eq:parameterization}
\vec{m}=\cos\theta\,\vec{e}_\text{\sc t}+\sin\theta\cos\phi\,\vec{e}_\text{\sc n}+\sin\theta\sin\phi\,\vec{e}_\text{\sc b},
\end{equation}
where $\vec{m}=\vec{M}/M_s$ is the normalized magnetization unit vector with $M_s$ being the saturation magnetization.

Magnetization dynamics of this system can be studied by means of phenomenological Landau-Lifshitz equations
\begin{equation}\label{eq:LLE}
-\sin\theta\dot\theta=\omega_0\frac{\delta\mathcal{E}}{\delta\mathcal{\phi}}+\alpha\sin^2\theta\dot{\phi},\quad\sin\theta\dot\phi=\omega_0\frac{\delta\mathcal{E}}{\delta\mathcal{\theta}}+\alpha\dot{\theta},
\end{equation}
where overdot indicates the time derivative, frequency $\omega_0=4\pi\gamma_0M_s$ determines the characteristic time scale of the system with $\gamma_0$ being the gyromagnetic ratio,  $\mathcal{E}=E/4\pi M_s^2$ is normalized total energy, and $\alpha$ is the damping coefficient.

We start with a simple model, which takes into account only two contributions to the total magnetic energy
\begin{equation} \label{eq:total-energy}
	\mathcal{E} =\mathcal{S}\int\limits_{-\infty}^{+\infty} \bigl[\ell^2\mathscr{E}_{\text{ex}}-k_{\text{t}}\cos^2\theta\bigr]\text{d}s,
\end{equation}
namely, exchange one $\mathscr{E}_{\text{ex}}$ and easy-tangential anisotropy -- the second term in (\ref{eq:total-energy}). Here $\mathcal{S}=\pi R^2$ is area of the wire cross-section, $\ell=\sqrt{A/4\pi M_s^2}$ is the exchange length with $A$ being exchange constant, and $k_{\text{t}}=K/(4\pi M_s^2)+1/4$ is the dimensionless anisotropy constant. Here $K>0$ is constant magnetocrystalline anisotropy of easy-tangential type and the term $1/4$ comes from the magnetostatic contribution. It is known\cite{Hillebrands06,Porter04,Slastikov12,Kravchuk14c} that for thin wires of circular and square cross-sections the magnetostatic energy is reduced to an effective easy-tangential shape anisotropy with $K_{\text{eff}}=\pi M_s^2$, including case of a curvilinear wire.\cite{Slastikov12} Competition of the exchange and anisotropy contributions results in the length scale of the system $w=\ell/\sqrt{k_{\text{t}}}$. For magnetically soft wires $K=0$ and $w=2\ell$. In (\ref{eq:total-energy}) and everywhere below we restrict ourselves with the case of thin wire $R\lesssim w$, thus we assume that the magnetization varies along the wire and it is uniform within a wire cross-section: $\vec{m}=\vec{m}(s)$. 


In terms of the angular parametrization (\ref{eq:parameterization}) the exchange energy density has a form\cite{Sheka15}
\begin{equation} \label{eq:Eex-density}
\mathscr{E}_{\text{ex}} =\left(\theta'+\kappa\cos\phi\right)^2+\left(\phi'\sin\theta-\kappa\cos\theta\sin\phi\right)^2,
\end{equation}
where the prime denotes derivative with respect to $s$. In (\ref{eq:Eex-density}) it is taken into account that a plane wire has zero torsion.

Let us first analyze static solutions of (\ref{eq:LLE}). In this case there is a solution $\phi=\phi_0=0,\,\pi$, which corresponds to a planar magnetization distribution: vectors $\vec{m}(s)$ lie within the wire plane. The corresponding function $\theta$ is determined by an inhomogeneous Sine-Gordon equation (for details see Appendix~\ref{app:stat-sol})
\begin{equation}\label{eq:SG}
w^2\theta''-\sin\theta\cos\theta=-w^2\kappa'\cos\phi_0.
\end{equation}
\begin{figure}
	\includegraphics[width=\columnwidth]{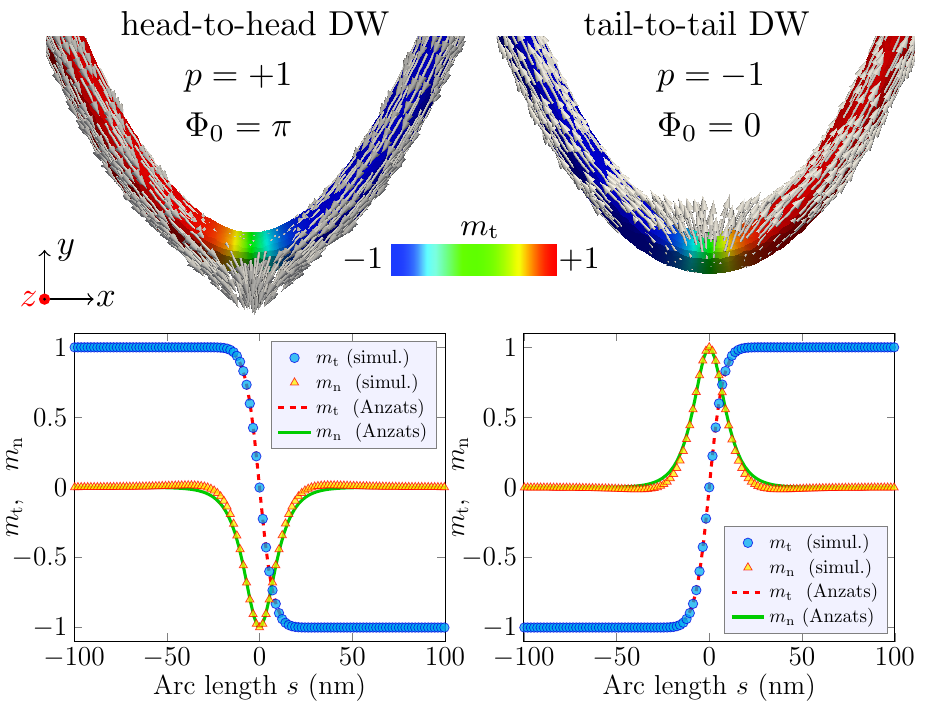}
	\caption{(Color online) Equilibrium state of the transversal domain wall at the  wire bend. Top row shows the magnetization distribution obtained using micromagnetic \texttt{NMAG} simulations for permalloy parabolic wire with $R=5$ nm, total length $L=1$~$\mu$m and $\kappa_0=0.05$ nm$^{-1}$ (only the bend vicinity is shown). The bottom row demonstrates comparison of magnetization components $m_{\text{t}}=\vec{m}\cdot\vec{e}_{\text{\sc t}}$ and $m_{\text{n}}=\vec{m}\cdot\vec{e}_{\text{\sc n}}$ obtained from the simulations (markers) and from the Ansatz~(\ref{eq:ansatz-stat}) (lines). In the latter case the domain wall width $\Delta=2\ell$ was used. }\label{fig:stat-DW}
\end{figure}
For the case $\kappa'\equiv0$ (rectilinear or circular wire) Eq.~(\ref{eq:SG}) has the well known domain wall solutions $\cos\theta=-p\tanh[(s-q)/\Delta]$ of head-to-head ($p=1$) or tail-to-tail ($p=-1$) types. Here $q$ determines the domain wall position and $\Delta=w$ is the wall width. For the case $\kappa'\not\equiv0$ an additional driving force appears and the curvature induced domain wall dynamics is expected. In the following, we consider a case of a localized curvature, i.e. $\kappa(\pm\infty)=0$ and $\kappa'(\pm\infty)=0$, see Figs.~\ref{fig:stat-DW},~\ref{fig:geometry}(a), this results in the boundary conditions $\cos\theta(\pm\infty)=\mp p$, the same as for a rectilinear wire. We also restrict ourselves with the case $w^2\kappa'\ll1$ assuming that the domain wall remains its form and the curvature influence results in a weak modification of the width $\Delta$. Therefore, to analyze the domain wall properties (static as well as dynamic ones) we use collective variable approach\cite{Slonczewski72,Thiele73} based on the simple $q-\Phi$ model\cite{Slonczewski72,Malozemoff79}

\begin{equation}\label{eq:ansatz-stat}
\cos\theta=-p\tanh\left[\frac{s-q(t)}{\Delta}\right], \quad \phi=\Phi(t),
\end{equation}
which is widely used for different types of domain walls and various drivings.\cite{Thiaville02a,Thiaville04,Thiaville05,Hillebrands06,Mougin07,Khvalkovskiy09,Landeros10,Thiaville12,Otalora12a,Otalora13}
The domain wall position $q$ and phase $\Phi$, which determines orientation of the transversal magnetization component, make a canonically conjugated pair of collective variables. The domain wall width $\Delta$ is shown to be a slaved variable,\cite{Hillebrands06,Landeros10} i.e. $\Delta(t)=\Delta[q(t),\Phi(t)]$. However, this is not the case when $\Phi$ depends on coordinate\cite{Kravchuk14} (it is possible for a three dimensional wire, when the torsion is present), in this case a generalization of the $q-\Phi$ model\cite{Kravchuk14} must be applied.  

Substituting now the Ansatz (\ref{eq:ansatz-stat}) into (\ref{eq:total-energy}) one obtains the total energy in the form (up to an additive constant)
\begin{equation}\label{eq:E-narrow-DW}
\frac{\mathcal{E}}{2\mathcal{S}}=\left(\frac{\ell^2}{\Delta}+ k_{\text{t}}\Delta\right)+\pi\ell^2p\cos\Phi\kappa(q)-\ell^2\Delta\sin^2\Phi\kappa^2(q),
\end{equation}
where the condition $\kappa\Delta\ll1$ was applied when integrating (\ref{eq:total-energy}), for details see Appendix~\ref{app:q-phi-det}. The first term in (\ref{eq:E-narrow-DW}) reflects the exchange-anisotropy competition which determines the domain wall width for a rectilinear wire. The second and the third terms originate from Dzyaloshinsky-like and anisotropy-like terms, respectively, which effectively appear in the exchange energy due to the curvature.\cite{Sheka15} Minimization of (\ref{eq:E-narrow-DW}) with respect to parameters $q$ and $\Phi$ results in the following equilibrium values $q_0$ and $\Phi_0$:
\begin{equation}\label{eq:equilibr-params}
\kappa'(q_0)=0,\quad \cos\Phi_0=-p.
\end{equation}
Thus, the domain wall is pinned at the maximum of the curvature, and the phase selectivity takes place: the head-to-head domain wall is always directed outward while the tail-to-tail one is directed inward the bend, see Fig.~\ref{fig:stat-DW}. The latter effect was recently observed experimentally.\cite{Kim14} There is an intuitive explanation: the choice of $\Phi_0$ defined by (\ref{eq:equilibr-params}) makes the magnetization distribution more homogeneous, see Fig.~\ref{fig:stat-DW}, which decreases the exchange energy.

As it follows from (\ref{eq:E-narrow-DW}), the equilibrium value of the domain wall width $\Delta_0=w$ is the same as for a rectilinear wire. However, if the wall is perturbed and the values of quantities ($q,\,\Phi$) deviate from the equilibrium (\ref{eq:equilibr-params}), the domain wall width is modified as follows $\Delta(q,\Phi)=\ell [k_{\text{t}}-k_{\text{b}}(q)\sin^2\Phi]^{-1/2}$, where $k_\mathrm{b}=\ell^2\kappa^2(q)$ is coefficient of the curvature induced effective easy-binormal anisotropy. The similar domain wall width modification caused by a geometrical constrain was discussed in Ref.~\onlinecite{Bruno99}.

As it follows from the Fig.~\ref{fig:stat-DW} the used Ansatz (\ref{eq:ansatz-stat}) and the obtained static results (\ref{eq:equilibr-params}) are in a good agreement with the simulations results. 


Let us now proceed to dynamical properties of the domain wall. In terms of the collective variables the equations of motion (\ref{eq:LLE}) take a form (see Appendix~\ref{app:q-phi-det})
\begin{equation}\label{eq:Dynamic-equations}
\dot q=\frac{\omega_0}{2\mathcal{S}}\frac{\partial\mathcal{E}}{\partial\Phi}+\alpha\Delta\dot{\Phi},\quad \dot \Phi=-\frac{\omega_0}{2\mathcal{S}}\frac{\partial\mathcal{E}}{\partial q}-\frac{\alpha}{\Delta}\dot{q}.
\end{equation}

Our goal is to study linear dynamics of the domain wall in vicinity of the equilibrium position.  With this purpose we introduce small deviations in the way $q(t)=q_0+\tilde{q}(t)$ and $\Phi(t)=\Phi_0+\tilde{\Phi}(t)$. For the limit case $\kappa\Delta_0\to0$ the equations of motion (\ref{eq:Dynamic-equations}) linearized with respect to the deviations read
\begin{equation}\label{eq:linearized}
(1+\alpha^2)\begin{Vmatrix}
\dot{\tilde{q}}\\\dot{\tilde{\Phi}}
\end{Vmatrix}\approx\omega_0\ell^2\pi
\begin{Vmatrix}
0&\kappa(q_0)\\
\kappa''(q_0)&-\alpha\frac{\kappa(q_0)}{\Delta_0}
\end{Vmatrix}\cdot\begin{Vmatrix}
\tilde{q}\\\tilde{\Phi}
\end{Vmatrix}.
\end{equation}
For the case of small $\alpha$ the solution of (\ref{eq:linearized}) results in harmonic decaying oscillations $\tilde{q}\propto\sin(\Omega t+\delta_0)e^{-\eta t}$, $\tilde{\Phi}\propto\cos(\Omega t+\delta_0)e^{-\eta t}$ with frequency 
\begin{equation}\label{eq:oscil-freq}
\Omega\approx\omega_0\ell^2\pi\sqrt{\kappa(q_0)|\kappa''(q_0)|}
\end{equation}
and modified effective friction
\begin{equation}\label{eq:friction}
\eta\approx\alpha\omega_0\frac{\pi}{2}\frac{\ell^2\kappa(q_0)}{\Delta_0}.
\end{equation}
The phase $\delta_0$ is determined by the initial conditions. In (\ref{eq:oscil-freq}) and (\ref{eq:friction}) it is taken into account that $\kappa(q_0)>0$ and $\kappa''(q_0)<0$ at the maximum point $q_0$.

It should be noted that an energy expression analogous to (\ref{eq:E-narrow-DW}) was earlier obtained for a certain case of a circular wire segment.\cite{Kruger07a} However, the circular geometry does not produce any geometrical pinning potential due to the constant curvature $\kappa\equiv\mathrm{const}$, which results in $\Omega=0$ and undefined $q_0$. Nevertheless, the domain wall can have an equilibrium position in circular segment in the presence of an external magnetic field.\cite{Kruger07a,Jamali11}

In more general case $\Delta_0\kappa\in(0,1)$ the Ansatz (\ref{eq:ansatz-stat}) leads to the following expressions for the eigen-frequency and effective friction
\begin{equation}\label{eq:oscil-freq-refined}
\begin{split}
&\Omega=\omega_0\frac{\ell^2}{\Delta_0^2}\sqrt{\left|\mathcal{AB}\right|},\qquad \eta=\alpha\omega_0\frac{\ell^2}{\Delta_0^2}\frac{|\mathcal{A}|+\mathcal{B}}{2},\\ &\mathcal{A}=\Delta_0^2\int\limits_{-\infty}^{+\infty}\!\!\!\frac{\kappa''(s)}{\cosh\frac{s-q_0}{\Delta_0}}\text{d}s,\\
&\mathcal{B}=\int\limits_{-\infty}^{+\infty}\!\!\!\frac{\kappa(s)}{\cosh\frac{s-q_0}{\Delta_0}}\text{d}s-\Delta_0\int\limits_{-\infty}^{+\infty}\!\!\!\frac{\kappa^2(s)}{\cosh^2\frac{s-q_0}{\Delta_0}}\text{d}s.
\end{split}
\end{equation}
In this case the equilibrium position $q_0$ is defined by the equation
\begin{equation}\label{eq:equilibr-q0}
\int\limits_{-\infty}^{+\infty}\!\!\!\frac{\kappa'(s)}{\cosh\frac{s-q_0}{\Delta_0}}\text{d}s=0
\end{equation}
and the equilibrium value of $\Phi_0$ coincides with one defined in (\ref{eq:equilibr-params}), for details see Appendix~\ref{app:q-phi-det}.

It is important to emphasize that the pinning potential and the corresponding domain wall oscillations have purely exchange origin, while the weak interactions, e.g.  magnetostatic one, contribute mediately via value of parameter $\Delta_0$.

\section{The case of parabolic bend}
\begin{figure*}
    \includegraphics[width=\textwidth]{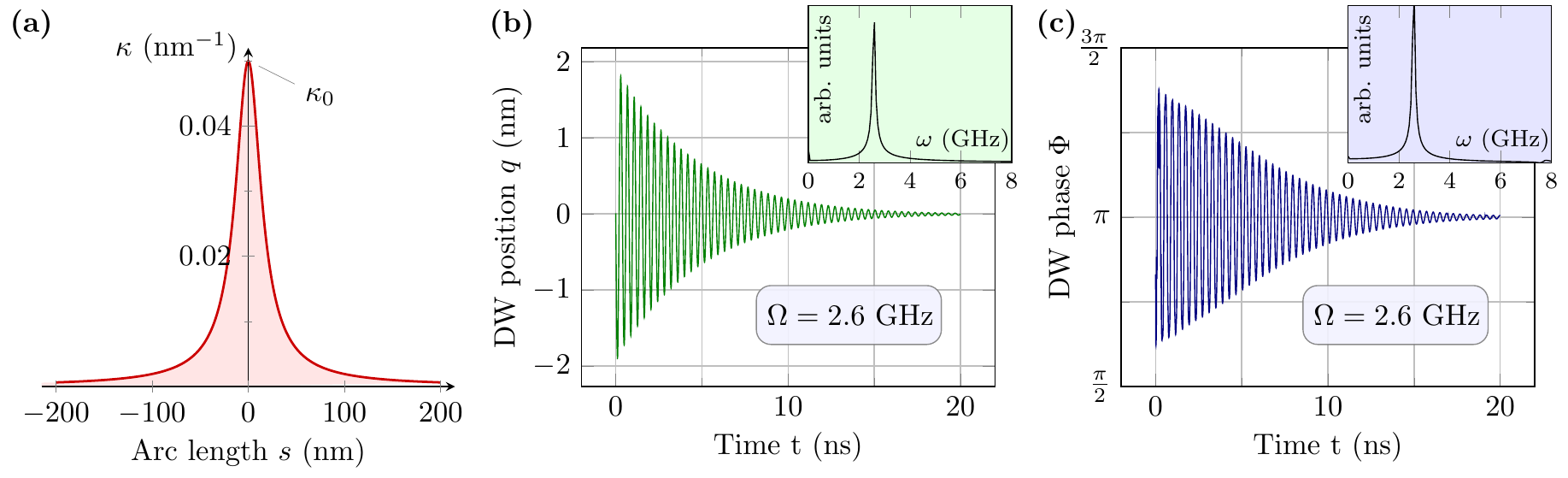}
	\caption{(Color online) (a) Curvature of the parabolic wire (\ref{eq:parabolic-gamma}) with $\kappa_0=0.05$ nm$^{-1}$. Insets (b) and (c) shows the time dependences (with the corresponding Fourier spectra) of position $q(t)$ and phase $\Phi(t)$ of the domain wall, perturbed at vicinity of the equilibrium point $q_0=0$. The data are obtained by means of micromagnetic simulations for a wire with $\kappa_0=0.05$~nm$^{-1}$ and $R=5$~nm.}
	\label{fig:geometry}
\end{figure*}

Let us now check the obtained general results for an example of parabolic wire, whose central line has a form
\begin{equation}\label{eq:parabolic-gamma}
\vec{\gamma}=x\hat{\vec x}+\kappa_0\frac{x^2}{2}\hat{\vec y}.
\end{equation}
The wire curvature $\kappa=\kappa_0(1+\kappa_0^2x^2)^{-3/2}$ as a function of natural parameter $s=\kappa_0^{-1}f(\kappa_0x)$, with $f(\xi)=\frac12\left[\xi\sqrt{1+\xi^2}+\mathrm{arcsinh}(\xi)\right]$, is shown in Fig.~\ref{fig:geometry}(a). It has a single maximum at point $s_0=0$ with extreme value $\kappa(s_0)=\kappa_0$. 

In accordance with (\ref{eq:equilibr-q0}) the equilibrium domain wall position is $q_0=0$, i.e. the domain wall is located at the extreme point of the bend. The equilibrium phase value $\Phi_0$ is determined by (\ref{eq:equilibr-params}), see Fig.~\ref{fig:stat-DW}.

In the narrow domain wall limit $\Delta_0\kappa_0\to0$ the eigenfrequency (\ref{eq:oscil-freq}) and effective friction read
\begin{equation}\label{eq:omega-narrow-parab}
\Omega\approx\omega_0\sqrt{3}\pi(\kappa_0\ell)^2,\qquad\eta\approx\alpha\omega_0\pi\ell\kappa_0/4.
\end{equation}

In order to verify our analytical results we performed two types of numerical simulation: (i) micromagnetic simulations of magnetically soft nanowire (\texttt{NMAG} code); (ii) numerical solution of the Landau-Lifshitz equations for a discrete chain of magnetic moments.
%

Let us start by considering the magnetically soft wires, whose geometries are determined by (\ref{eq:wire_definition}) and (\ref{eq:parabolic-gamma}). The radius of cross-section $R=5$ nm and length $L=1$~$\mu$m are fixed for all studied samples, while the extreme curvature $\kappa_0$ is varied in the range $\kappa_0\in[0.005,\,0.05]$~nm$^{-1}$ with step $\Delta\kappa_0=0.005$~nm$^{-1}$. 

Permalloy is chosen as a material with the following parameters: exchange constant $A=13$~pJ/m, saturation magnetization $M_s=860$ kA/m, and damping coefficient $\alpha=0.01$. These parameters results in the exchange length $\ell\approx3.7$ nm and $\omega_0=30.3$~GHz. Thermal effects and anisotropy are neglected.

The magnetization dynamics is studied by means of numerical simulation of the Landau-Lifshitz equation applying the \texttt{NMAG} code~\cite{Fischbacher07}, wherein an irregular tetrahedral mesh with cell size about 1.75~nm is used. Only three magnetic interactions were taken into account, namely exchange, magnetostatic and Zeeman contributions.

The numerical experiment consists of three steps. Initially, we relax the domain wall structure in an over-damped regime~($\alpha=0.1$) in order to determine the equilibrium values of collective variables: position $q_0$ and phase $\Phi_0$. The obtained results fully coincides with the prediction (\ref{eq:equilibr-params}). To determine the values of $q$ and $\Phi$  we extract the curvilinear magnetization components $m_{\text{t}}=\vec{m}\cdot\vec{e}_{\text{\sc t}}$, $m_{\text{n}}=\vec{m}\cdot\vec{e}_{\text{\sc n}}$, and $m_{\text{b}}=\vec{m}\cdot\vec{e}_{\text{\sc b}}$ from the simulation data, and apply fitting with the Ansatz (\ref{eq:ansatz-stat}).  Namely, the position $q$ is determined as a fitting parameter for the function $m_{\text{t}}(s)=-p\tanh[(s-q)/\Delta]$, then the phase is determined from the equation $\tan\Phi=m_{\text{b}}(q)/m_{\text{n}}(q)$.

\begin{figure}
\includegraphics[width=\columnwidth]{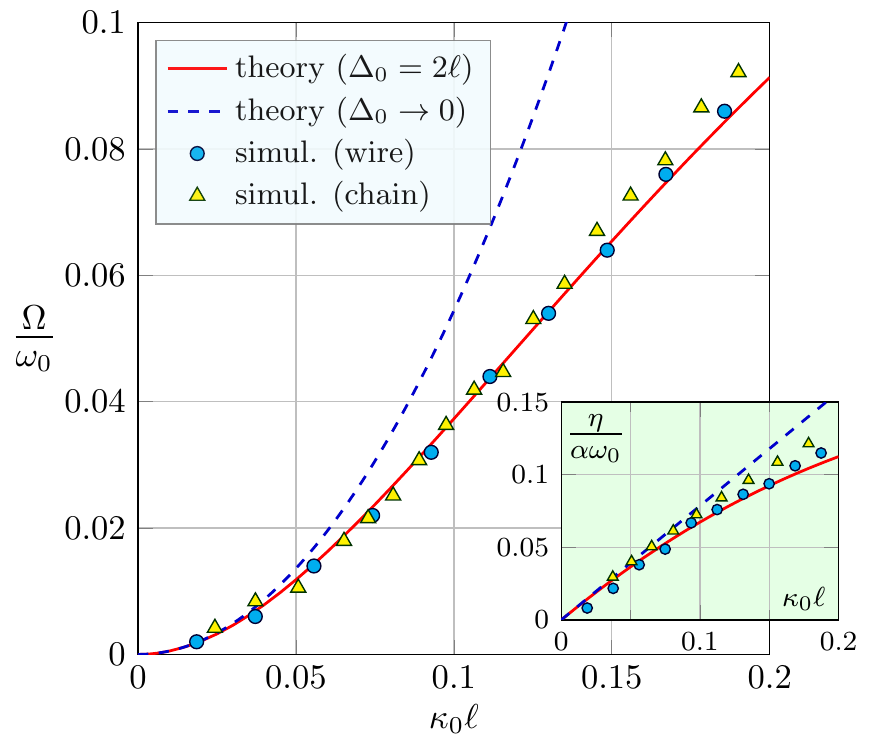}\\
\caption{(Color online) Eigenfrequency of the domain wall oscillations in vicinity of the equilibrium -- the extreme point of parabolic bend, see Fig.~\ref{fig:stat-DW}. Solid and dashed lines correspond to the predictions (\ref{eq:oscil-freq-refined}) and its narrow-wall asymptotics (\ref{eq:omega-narrow-parab}), respectively. Markers shows the results of numerical simulations for nanowires (disks) and discrete chains of magnetic moments (triangles). The inset demonstrates dependence of the effective friction on the curvature parameter.}
\label{fig:wall_dynamics}
\end{figure}

In the second step we slightly perturb the domain wall phase $\Phi$ by applying a weak magnetic field $\vec B=B_0\vec{e}_z$ perpendicularly to the wire plane, where $B_0=25$~mT. After the system relaxation in the applied field $\vec{B}$ the field is switched off and the magnetization dynamics is simulated for the natural value of the damping coefficient (the third step). Since the variables $\Phi$ and $q$ are canonically conjugated, see Eqs.~(\ref{eq:Dynamic-equations}), the perturbed dynamics of $\Phi$ induces the dynamics of $q$, i.e. the domain wall starts to move.  Typical time dependences of these collective variables are shown in Fig.~\ref{fig:geometry}(b,c), where one can see harmonic decaying oscillations with well pronounced frequency $\Omega$ and friction parameter $\eta$. In this way we determine $\Omega$ and $\eta$ for all range of the studied curvatures $\kappa_0$. The frequency $\Omega$ as well as the friction $\eta$ increases with the curvature increasing, see Fig.~\ref{fig:wall_dynamics} (round markers). One should note a good agreement of the obtained numerical results with the analytical prediction (\ref{eq:oscil-freq-refined}) with $\Delta_0=2\ell$, see solid lines in Fig.~\ref{fig:wall_dynamics} and Appendix~\ref{app:q-phi-det} for details. This means that the approximation of magnetostatic interaction by the effective easy-tangential anisotropy is physically sound for a domain wall dynamics in the thin nanowires.

Additionally, we study dynamics of discrete chains of magnetic moments $\vec{m_i}$, with $i =\overline{1,N}$, aligned along parabolic line (\ref{eq:parabolic-gamma}). The moments are aligned equidistantly with the fixed step size $\Delta s$ along the chain.  As previously, only three magnetic interactions are taken into account, namely exchange (with $\ell=3\Delta s$), dipole-dipole and Zeeman contribution. Total number of moments is fixed $N=76$ while the curvature parameter $\kappa_0$ is varied but the restriction $\kappa_0\Delta s\ll1$. Dynamics of this system is described by a set of $N$ vector Landau-Lifshitz equations which a solved by means of common numerical methods, in details this procedure is described in Ref.~\onlinecite{Sheka15b}.

The domain wall oscillations are studied in the same way as for the wires described above. The resulting eigenfrequencies and friction parameters are shown in  Fig.~\ref{fig:wall_dynamics} by triangles. 

In conclusion, we demonstrate an effect of curvature induced pinning of a transversal head-to-head (tail-to-tail) domain wall at a local wire bend. The curvature induced pinning potential has purely exchange origin. We obtain an expression (\ref{eq:oscil-freq-refined}) for frequency and effective friction of the free domain wall oscillations within the pinning potential, and we demonstrate its validity numerically for the range of parameters $\kappa\Delta\sim10^{-1}$. For the case of a narrow domain wall or weak curvature ($\kappa\Delta\sim10^{-2}$) the approximations (\ref{eq:oscil-freq}) and (\ref{eq:friction}) can be used. The good agreement of analytical predictions and results of full scale numerical simulations for magnetically soft cases ($K=0$) demonstrates that the approximation of magnetostatic interaction by the effective easy-tangential anisotropy is physically sound for a domain wall dynamics in the thin nanowires.

\section{Acknowledgments}
The present work was partially supported by the Program of Fundamental Research of the Department of Physics and Astronomy of the National Academy of Sciences of Ukraine (project No. 0112U000056).

\appendix
\section{Static solutions for a planar wire}\label{app:stat-sol}

Static form of Eqs.~(\ref{eq:LLE}) reads $\delta\mathcal{E}/\delta\theta=0$, $\delta\mathcal{E}/\delta\phi=0$. Taking into account form of energy (\ref{eq:total-energy}) and exchange energy density (\ref{eq:Eex-density}) one obtains the following set of equations
\begin{subequations}\label{eq:stat}
\begin{align}\label{eq:stat-theta}
	\theta''-&\sin\theta\cos\theta\left[\phi'^2-\kappa^2\sin^2\phi+w^{-2}\right]\\ \nonumber
	-&2\phi'\kappa\sin^2\theta+\kappa'\cos\phi=0,
\end{align}	
\begin{align}\label{eq:stat-phi}
	\phi''+&2\theta'\left[\cot\theta\phi'+\kappa\sin\phi\right]+\kappa^2\sin\phi\cos\phi\\ \nonumber
	-&\kappa'\cot\theta\sin\phi=0.
\end{align}	
\end{subequations}
Equation (\ref{eq:stat-phi}) has solutions $\phi=\phi_0=0,\,\pi$. Substitution of this solution into (\ref{eq:stat-theta}) results in equation (\ref{eq:SG}), which determines the function $\theta(s)$.

\section{Details of the $q-\Phi$ model usage}
\label{app:q-phi-det}
The equations of motion (\ref{eq:LLE}) are the Euler-Lagrange equations
\begin{equation}\label{eq:Euler-Lagrange}
\frac{\delta\mathcal{L}}{\delta\xi_i}-\frac{\mathrm d}{\mathrm d t}\frac{\delta\mathcal{L}}{\delta\dot{\xi}_i}=\frac{\delta\mathcal{F}}{\delta\dot{\xi}_i},\qquad \xi_i=\theta,\,\phi
\end{equation}
for Lagrange function\cite{Doering48}
\begin{equation}\label{eq:Lagrange}
\mathcal{L}=-\frac{\mathcal{S}}{\omega_0}\int\limits_{-\infty}^{\infty}\phi\sin\theta\dot{\theta}\mathrm{d}s-\mathcal{E}
\end{equation}
and dissipative function
\begin{equation}\label{eq:diss-func}
\mathcal{F}=\frac{\alpha}{2}\mathcal{S}\int\limits_{-\infty}^{\infty}\left[\dot{\theta}^2+\sin^2\theta\dot{\phi}^2\right]\mathrm{d}s.
\end{equation}
Substituting the Ansatz (\ref{eq:ansatz-stat}) into (\ref{eq:Lagrange}) and (\ref{eq:diss-func}) and performing the integration over $s$ (along the curve)  one obtains the effective Lagrange and effective dissipative functions in the form
\begin{equation}\label{eq:L-F-in-CV}
\mathcal{L}_{\text{eff}}=\frac{2\mathcal{S}}{\omega_0}\Phi\dot{q}-\mathcal{E},\quad \mathcal{F}_{\text{eff}}=\alpha\frac{\mathcal{S}}{\omega_0}\left[\frac{\dot{q}^2}{\Delta}+\Delta\dot{\Phi}^2\right].
\end{equation}
Substituting (\ref{eq:L-F-in-CV}) into the Euler-Lagrange equations (\ref{eq:Euler-Lagrange}) one obtains the equations of motion (\ref{eq:Dynamic-equations}) written in terms of the collective variables.

Let us now consider energy of the system $\mathcal{E}$. Substituting the Ansatz (\ref{eq:ansatz-stat}) into the energy expression (\ref{eq:total-energy}) and performing the integration one obtains (up to an additive constant)
\begin{equation}\label{eq:E-CV}
\begin{split}
&\frac{\mathcal{E}}{2\mathcal{S}}=\frac{\ell^2}{\Delta}+\Delta k_t+\frac{\ell^2}{\Delta^2}\left[pF_1\cos\Phi-\frac{F_2}{2}\sin^2\Phi\right]\\
&F_n=\int\limits_{-\infty}^{\infty}f^n(s)\mathrm{d}s,\qquad f(s)=\frac{\kappa(s)\Delta}{\cosh\frac{s-q}{\Delta}}.
\end{split}
\end{equation}

In the limit case $\kappa\Delta\to0$ the expression (\ref{eq:E-CV}) is reduced to (\ref{eq:E-narrow-DW}). For the more general case $0<\Delta\kappa<1$ one has $f<1$ and consequently $F_1>F_2$. In this case the value of phase $\Phi_0$ which minimizes the energy (\ref{eq:E-CV}) is determined as $\cos\Phi_0=-p$ and the corresponding value of the equilibrium domain wall position $q_0$ is determined by the equation (\ref{eq:equilibr-q0}). The case $\Delta\kappa>1$ (domain wall width is larger than the curvature radius) is not considered here, because the application of the Ansatz (\ref{eq:ansatz-stat}) is questionable in this case and a precise solution of equation (\ref{eq:SG}) is required.

The equations of motion linearized in vicinity of the equilibrium $q=q_0$, $\Phi=\Phi_0$ read
 
 \begin{equation}\label{eq:linearized-gen}
 (1+\alpha^2)\begin{Vmatrix}
 \dot{\tilde{q}}\\\dot{\tilde{\Phi}}
 \end{Vmatrix}=\omega_0\frac{\ell^2}{\Delta_0^2}
 \begin{Vmatrix}
 \alpha\mathcal{A}&\Delta_0\mathcal{B}\\
 \frac{\mathcal{A}}{\Delta_0}&-\alpha\mathcal{B}
 \end{Vmatrix}\cdot\begin{Vmatrix}
 \tilde{q}\\\tilde{\Phi}
 \end{Vmatrix},
 \end{equation}
where $\tilde{q}=q-q_0$ and $\tilde{\Phi}=\Phi-\Phi_0$ are small deviations from the equilibrium and quantities $\mathcal{A}$ and $\mathcal{B}$ are defined in (\ref{eq:oscil-freq-refined}). For the case of small $\alpha$ the solution of (\ref{eq:linearized-gen}) results in harmonic decaying oscillations $\tilde{q}\propto\sin(\Omega t+\delta_0)e^{-\eta t}$, $\tilde{\Phi}\propto\cos(\Omega t+\delta_0)e^{-\eta t}$ with frequency $\Omega$ and effective friction $\eta$ presented in (\ref{eq:oscil-freq-refined}).

For the case of parabolic wire (\ref{eq:parabolic-gamma}) one obtains $\mathcal{A}=\mathcal{A}(\Delta_0\kappa_0)$ and $\mathcal{B}=\mathcal{B}(\Delta_0\kappa_0)$, where
\begin{equation}\label{eq:AB-parapol}
\begin{split}
&\mathcal{A}(x)=3x^2\int\limits_{-\infty}^{+\infty}\frac{(5\xi^2-1)\mathrm{sech}\frac{f(\xi)}{x}}{(1+\xi^2)^4}\mathrm{d}\xi,\\
&\mathcal{B}(x)=\int\limits_{-\infty}^{\infty}\left[1-x\frac{\mathrm{sech}\frac{f(\xi)}{x}}{(1+\xi^2)^{\frac{3}{2}}}\right]\frac{\mathrm{sech}\frac{f(\xi)}{x}}{1+\xi^2}\mathrm{d}\xi.
\end{split}
\end{equation}
These expressions are used to calculate the eigenfrequencies and friction parameters in (\ref{eq:oscil-freq-refined}), see solid lines in Fig.~\ref{fig:wall_dynamics}.

%

%

\end{document}